\documentstyle[osa,aps,prl,epsfig]{revtex}
\begin{document}
\twocolumn[\hsize\textwidth\columnwidth\hsize\csname
@twocolumnfalse\endcsname

%\documentclass[twocolumn,showpacs,preprintnumbers,amsmath,amssymb]{revtex4}
%\documentclass[preprint,eqsecnum,aps,twocolumn,twoside]{revtex4}
%\documentclass[preprint,showpacs,eqsecnum,aps]{revtex4}
%\documentstyle[eqsecnum,aps]{revtex4}
%\def\btt#1{{\tt$\backslash$#1}}
%\def\BibTeX{\rm B{\sc ib}\TeX}
%\renewcommand {\theequation}{\arabic{equation}}
%\documentclass [12pt]{article}
%\begin{document}
%\draft

\title{Temperature dependence and control of the Mott transition in VO$_2$-based devices}
\author{Hyun-Tak Kim (htkim@etri.re.kr), B. G. Chae, D. H. Youn, S. L. Maeng, and K. Y. Kang}
\address{Telecom. Basic Research Lab, ETRI, Daejeon 305-350, Korea}
\date{December 3, 2003}
\maketitle

\begin{abstract}
The transition voltage of an abrupt metal-insulator transition
(MIT), observed by applying an electric field to two-terminal
devices fabricated on a Mott insulator VO$_{2}$ film, decreases
with increasing temperature up to 334K. The abrupt current jump
disappears above 334 K near the MIT temperature. These results
suggest that the mechanism of the abrupt MIT induced by
temperature is the same as that by an electric field. The
magnitude of the current jump (a large current) decreases with
increasing external resistance; this is an important observation
in terms of applying the abrupt MIT to device applications.
Furthermore, the temperature and resistance dependence of the MIT
cannot be explained by the dielectric breakdown although a current
jump known as breakdown is similar to that observed in an abrupt
MIT.\\ \\ \\
%Corresponding author: Hyun-Tak Kim,  ~~~~htkim@etri.re.kr
\end{abstract}
%\pacs {71.27.+a, 71.30.+h}
]

%\narrowtext

Thin films of vanadium dioxide, VO$_2$, have been extensively
studied for electronic and electro-optic device applications
\cite{Stefanovich,Chudnovskiy,kimTR} of an abrupt first-order
metal-insulator transition (MIT) at a critical temperature
$T_c\approx$340 K \cite{Natale-13,Borek-14,Muraoka}. Recently, a
new type of abrupt MIT, having an abrupt jump of driving current
in a two-terminal device fabricated on an epitaxial VO$_{2}$ film,
has been demonstrated by electric field excitation \cite{HTKim-8}.
The abrupt MIT does not undergo a structural phase transition
\cite{Lim}, as predicted by Mott for an abrupt first-order MIT
driven by strongly correlated electronic Coulomb energy
\cite{Mott-1}.

When an abrupt MIT occurs and excess current flows in a device,
the device can be damaged or its characteristics can be degraded.
Furthermore, the electric I-V characteristics of the abrupt MIT
are similar to those of dielectric breakdown observed at high
electric fields in thin AlO$_x$ and HfO$_2$ gate insulators
\cite{Bhalla,Park}. Clearer evidence of the Mott transition, a
control method of the excess current for device applications
utilizing abrupt MITs, and evidence of the difference between a
breakdown and an abrupt MIT are important unresolved issues in
this field.

In this paper, we measure the temperature dependence of abrupt
MITs in VO$_{2}$ driven by a DC electric field to suggest evidence
of the Mott transition, and control the magnitude of the abrupt
current jump (or excess current) using an external resistance. The
effect of measurement on the magnitude of the observed current is
briefly discussed. An important difference between a breakdown and
an abrupt MIT is also given.

Thin films of VO$_{2}$ have been deposited on (1102)
Al$_{2}$O$_{3}$ and Si substrates by laser ablation \cite{Youn}.
The thickness of the VO$_{2}$ films is about 900 $\textrm{\AA}$.
For two-terminal devices, Ohmic-contacted Au/Cr electrodes on
VO$_{2}$ films with a channel width of 25 ${\mu}m$ and a channel
length of 5 ${\mu}m$ were patterned by photo-lithography and
lift-off. I-V characteristics of the devices were measured by a
precision semiconductor parameter analyzer (HP4156B).

Figure 1 (a) shows the temperature dependence of the resistance of
an epitaxial VO$_{2}$ film I. The resistance

\begin{figure}
\vspace{0.0cm}
\centerline{\epsfysize=10cm\epsfxsize=8cm\epsfbox{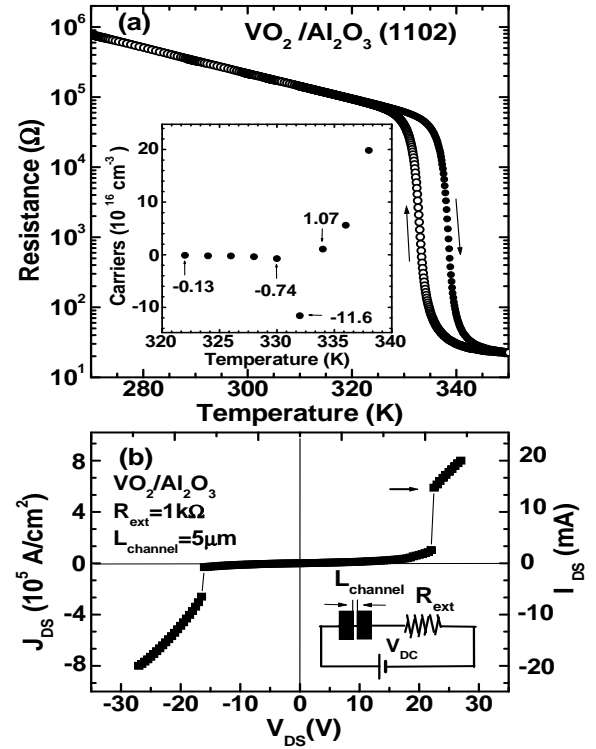}}
\vspace{0.5cm} \caption{(a) Temperature dependence of the
resistance of a VO$_{2}$ film I. Abrupt MITs are clearly shown.
The inset in (a) denotes the number of carriers obtained by Hall
measurements. A change of carriers from holes to electrons is
shown at 332 K. The minus sign indicates that the carriers are
holes. (b) DC electric field dependence of current density
measured by a two-terminal device I patterned on a VO$_2$ film
II/Al$_2$O$_3$. The inset in (b) denotes a circuit of a two
terminal device.}
\end{figure}

decreases with increasing temperature and shows an abrupt MIT at a
critical temperature $T_{c}\approx$340 K (68$^{\circ}$C). This is
consistent with previous measurements \cite{Natale-13,Borek-14}.
It was proposed that this abrupt MIT is due to the structural
phase transition from monoclinic below $T_{c}$ to tetragonal above
$T_{c}$ \cite{Morin-15,McWhan-20}. The decrease of the resistance
up to 340 K indicates an increase of hole carriers, and two kinds
of electron and hole carriers coexist near $T_{c}\approx$340 K, as
shown in the inset of Fig. 1 (a). Owing to a mixing of electrons
and holes, the number of carriers at temperatures from 332 to 340
K is not exactly determined. The number of hole carriers at
$T_{c}\approx$340 K is expected to be $n_c~\approx$ 3 $\times$
10$^{18}$ \textit{cm$^{-3}$} from the Mott criterion
\cite{Stefanovich,Mott-1}, based on an exponential decrease of
resistance (increase of carrier) with increasing temperature.
$n_c$ corresponds to 0.018\% of $d$-band charges. In the metal
regime above 340 K, the major carriers are electrons, as shown in
the inset of Fig. 1 (a). Fig. 1 (b) shows the drain-source-voltage
dependence of conducting current, $I_{DS}$, density, $J_{DS}$, of
the flow between two terminals (drain-source) for a VO$_{2}$ film
II. An abrupt current jump near the transition voltage $V_{t}$ =
20 $V$ is shown and Ohmic behavior as a characteristic of metal is
also exhibited over the transition voltage. This is a typical
characteristic of a first-order transition and is reproducible
more than 1,500 times.

%\newpage
\begin{figure}
\vspace{-.5cm}
\centerline{\epsfysize=6cm\epsfxsize=8.0cm\epsfbox{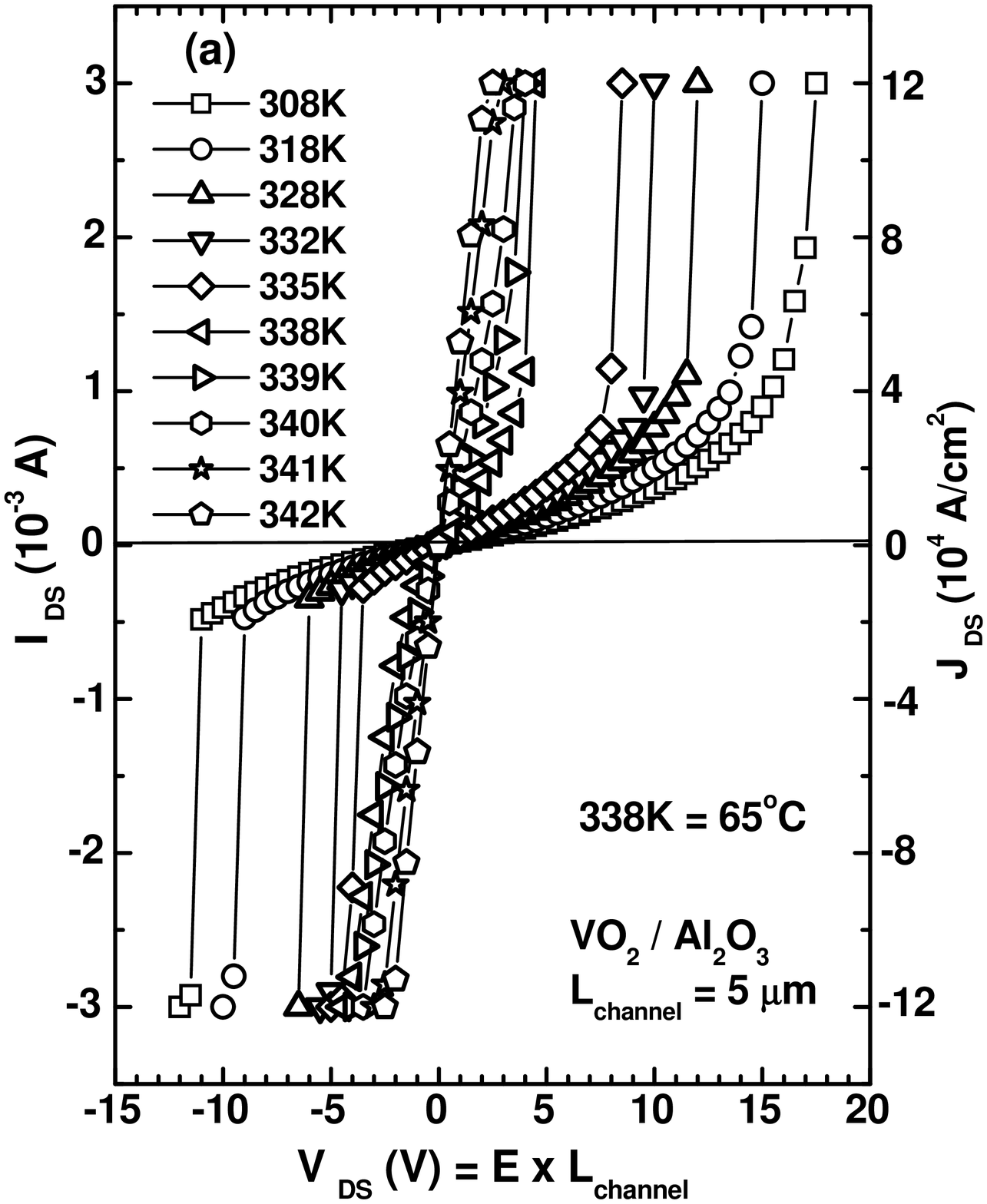}}
\vspace{-1.0cm}
\centerline{\epsfysize=6cm\epsfxsize=8.0cm\epsfbox{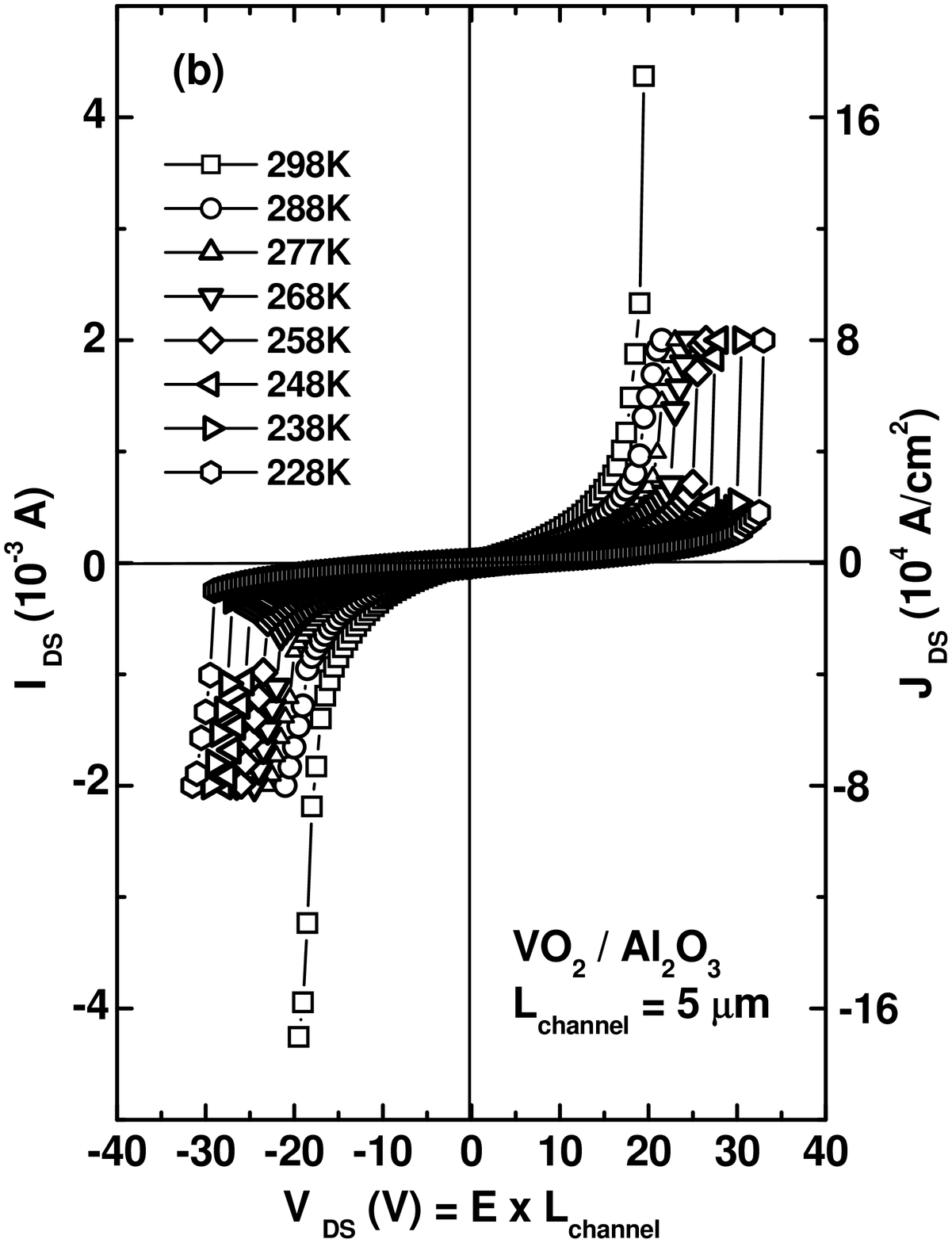}}
\vspace{0.5cm} \caption{Temperature dependence of abrupt MIT
observed at device II fabricated on a VO$_2$ film III/Al$_2$O$_3$
substrate above (Fig. a) and below (Fig. b) room temperature.
Above $T$ = 66$^{\circ}$C, apparent metallic behavior appears.}
\end{figure}

Figures 2 (a) and (b) show the temperature dependence of the
abrupt MIT measured at device II fabricated on a
VO$_2$/Al$_2$O$_3$ film III. The transition voltage of the abrupt
MIT decreases with increasing temperature. This arises from
excitation of hole charges by temperature. At 338 K, near the
transition temperature of the abrupt MIT, and beyond, $I_{DS}$s
follow Ohmic behavior without any current jump, in contrast to the
MITs with a current jump below 338 K; this is observed for the
first time. Note that the device was protected by a compliance
current of 3 mA and the measurement was carried out without
external resistance.

\begin{figure}
\vspace{-0.5cm}
\centerline{\epsfysize=6cm\epsfxsize=8cm\epsfbox{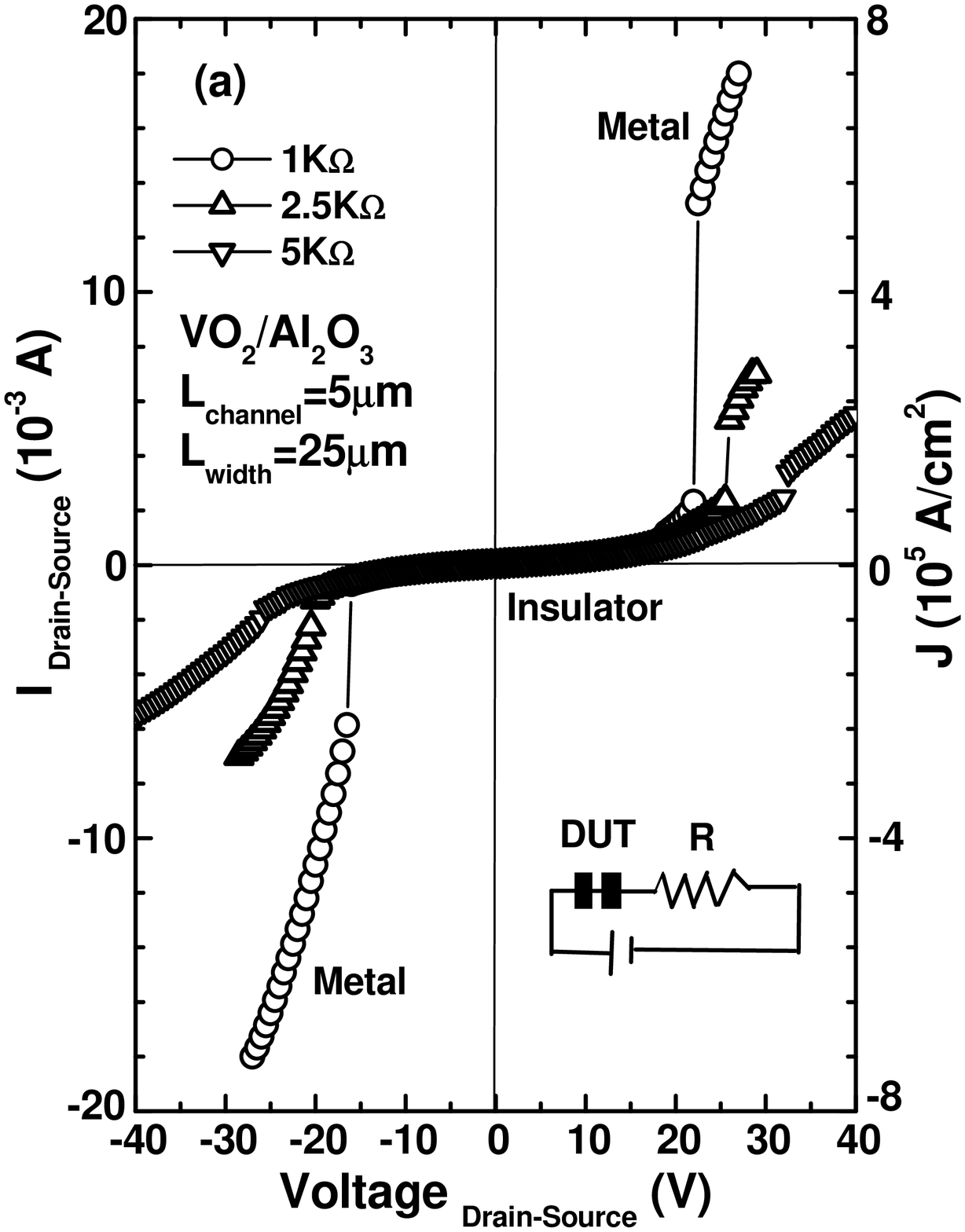}}
\vspace{-1.0cm}
\centerline{\epsfysize=6cm\epsfxsize=8cm\epsfbox{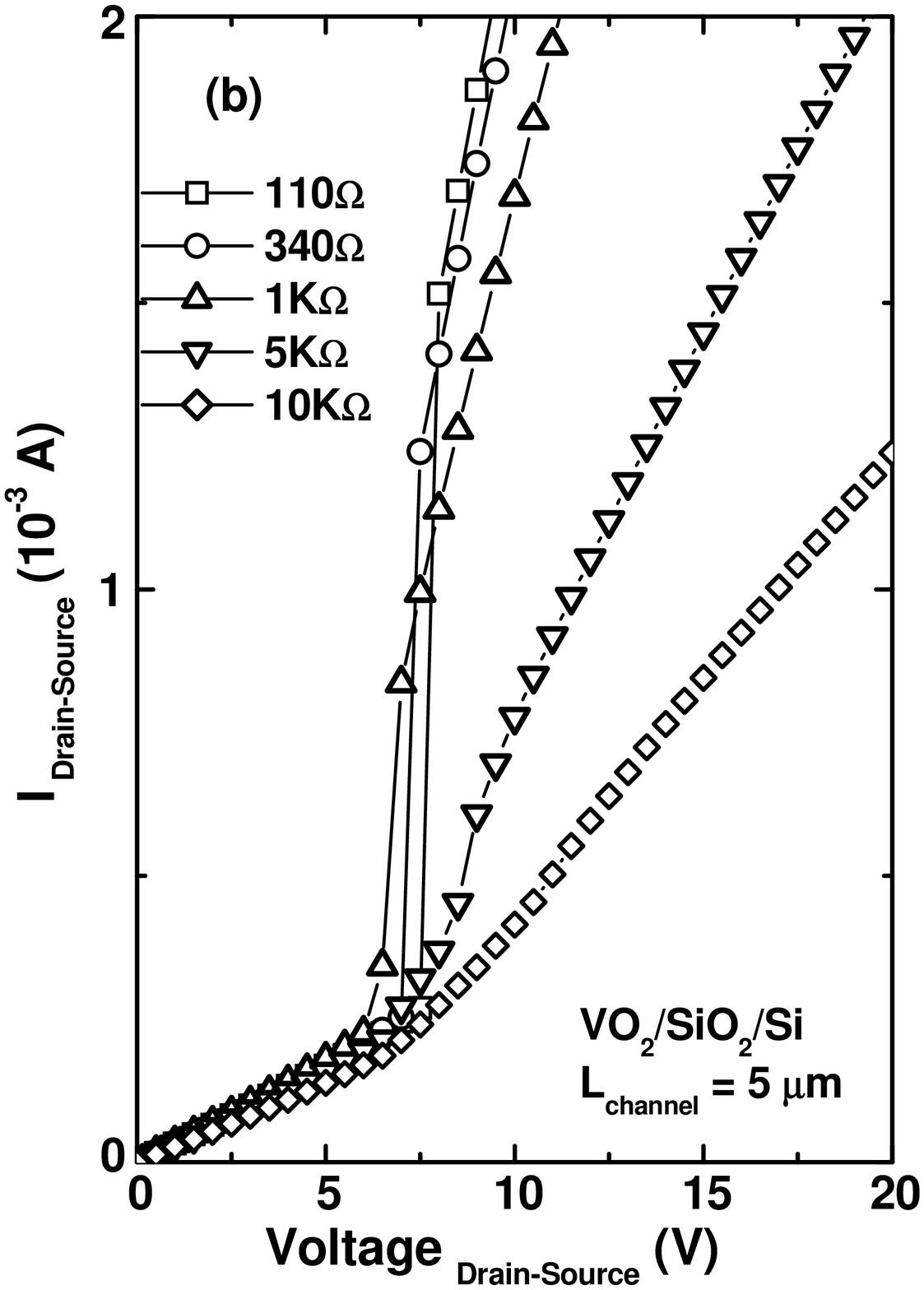}}
\vspace{0.0cm} \caption{External resistance dependence of abrupt
MIT measured,  (Fig. a) at device I fabricated on a VO$_2$ film
II/Al$_2$O$_3$ substrate, (Fig. b) at device III fabricated on a
VO$_2$ film IV/SiO$_2$/Si substrate.}
\end{figure}

The temperature dependence also provides decisive information for
revealing the mechanism of the abrupt jump. When the number of
total holes, $n_{tot}$, in the hole levels is given by
$n_{tot}=n_b + n_{free}(T,E)$, where $n_b$ is the number of bound
holes in the levels and $n_{free}(T,E)$ is the number of holes
freed by temperature, $T$, and electric field, $E$, from the
levels, $n_b$ decreases with increasing $n_{free}(T,E)$. For the
abrupt jump, ${\triangle}n{\equiv}n_c-n_{free}$=0 should be
satisfied, where $n_c~{\approx}~3{\times}10^{18}~cm^{-3}$, as
predicted by Mott \cite{Stefanovich,Mott-1}. At $T_{c}\approx$340
K, it is suggested, as decisive evidence of the Mott transition,
that the abrupt current jump will disappear, because
$n_{free}(T{\approx}340K, E{\approx}0)$=$n_c$
($i.e.~{\triangle}n$=0) is excited by only temperature, as shown
in Fig. 2(a). Below $T_{c}\approx$340 K, the abrupt MIT voltage
decreases with increasing temperature, because, from
$n_c{\equiv}n_{free}(T,E)$=$n_{free}(T)$ + $n_{free}(E)$, the
increase of $n_{free}(T)$ with increasing temperature decreases
$n_{free}(E)$. Thus, it is revealed that the mechanism of the
abrupt MIT excited by temperature (Fig. 1) is the same as that by
an electric field. In addition, if the abrupt current jump occurs
by breakdown due to a high field, the temperature dependence and
the change near 338 K of MIT cannot be explained.

Figures 3 (a) and (b) show the external resistance dependence of
the magnitude of abrupt current jumps observed at device II
fabricated on a VO$_2$/Al$_2$O$_3$ film II and device III
fabricated on a VO$_2$/SiO$_2$/Si film III, respectively. With
increasing external resistance, the magnitude of the abrupt
current jump decreases and the MIT voltage increases. This can be
explained by our model, as shown in Fig. 4 (a), where the metal
region decreases with increasing external resistance in the
measurement region. If only the metal region in Fig. 4 (a) is
measured, the magnitude of the current jump might be of an order
of $\sim$10$^7$ A/cm$^2$, the current density of a good metal.
That is, $I_{DS}$s observed at 5 K$\Omega$ do not have the
characteristics of a current jump, while, $I_{DS}$s measured at 1
K$\Omega$ and less resistance, display jumps, as shown in Figs.
3(a) and (b). This indicates that the observed $I_{DS}$ changes
with an external resistance change, even though the intrinsic
metal characteristic remain unchanged. Thus, the observed current
density, $J_{DS}$, of an order of $\sim$10$^5$ A/cm$^2$ in Fig. 3
(a) is an average of the metal region over the measurement region,
as shown in Fig. 4 (b). The average is the effect of measurement.
Furthermore, since the VO$_2$ film has electrons and holes, as
observed by Hall measurement in Fig. 1, it is regarded that the
VO$_2$ film is intrinsically inhomogeneous, although an external
resistance effect is excluded. The inhomogeneity is an intrinsic
characteristic of a material with an abrupt current jump and was
confirmed by high resolution cross-sectional transmission-electron
microscopy \cite{Youn}. Thus, a true current jump cannot be
measured in an inhomogeneous system, as shown in Fig. 4, as has
been explained by the extended Brinkman-Rice picture \cite{Kim-7}.

\begin{figure}
\vspace{-0.7cm}
\centerline{\epsfysize=6cm\epsfxsize=7cm\epsfbox{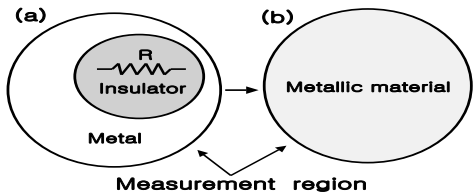}}
\vspace{-2.0cm} \caption{When a device is measured, the resistance
is included in the measurement region. This indicates that the
measurement region is inhomogeneous.}
\end{figure}

We have observed a true dielectric breakdown which forms a current
path with a very low source-drain voltage (or field) and a high
current such as a 'short' phenomenon, after a device is applied to
a very high current or electric field. After the dielectric
breakdown occurs, the current jump is not observed. The dielectric
breakdown is regarded as a device breakdown. The abrupt current
jump is reproducible without degradation even in thousands of
measurements, when an external resistance is attached in the
circuit. The external resistance is a decisive key to controlling
the abrupt MIT for application devices.

In conclusion, the temperature dependence of the abrupt MIT
explains the mechanism of the abrupt Mott MIT. In particular, the
external resistance dependence of the magnitude of the current
jump is an important key to application of the abrupt MIT to optic
and electronic devices. Furthermore, the breakdown phenomenon
observed in thin gate insulators \cite{Bhalla,Park} may be the
abrupt Mott MIT.

%\newpage

\end{document}